\pdfoutput=1

\documentclass[acmtog, authorversion]{acmart}

\AtBeginDocument{%
  \providecommand\BibTeX{{%
    \normalfont B\kern-0.5em{\scshape i\kern-0.25em b}\kern-0.8em\TeX}}}

\setcopyright{acmcopyright}
\copyrightyear{2023}
\acmYear{2023}

\acmConference[EAAMO'23]{Make sure to enter the correct
  conference title from your rights confirmation email}{October}{2023}
\acmBooktitle{EAAMO'23} 
\acmPrice{15.00}
\acmISBN{XXXXXXX.XXXXXXX}

\usepackage{xcolor}

\begin{document}

\title{PaperCard for Reporting Machine Assistance in Academic Writing}


\author{Won Ik Cho}
\authornote{Both authors contributed equally to this research. \textsuperscript{$\dagger$}Work done after graduation.}
\affiliation{%
  \institution{Seoul National University\textsuperscript{$\dagger$}}
  \city{Seoul}
  \country{South Korea}
}
\email{tsatsuki6@gmail.com}

\author{Eunjung Cho}
\authornotemark[1]
\affiliation{%
  \institution{ETH Zürich}
  \city{Zürich}
  \country{Switzerland}
}
\email{ejcho917@gmail.com}

\author{Kyunghyun Cho}
\affiliation{%
  \institution{New York University and Genentech}
  \city{New York}
  \country{USA}}
\email{kyunghyun.cho@nyu.edu}

\renewcommand{\shortauthors}{Cho, Cho, and Cho}

\begin{abstract}
   Academic writing process has benefited from various technological developments over the years including search engines, automatic translators, and editing tools that review grammar and spelling mistakes. They have enabled human writers to become more efficient in writing academic papers, for example by helping with finding relevant literature more effectively and polishing texts. While these developments have so far played a relatively assistive role, recent advances in large-scale language models (LLMs) have enabled LLMs to play a more major role in the writing process, such as coming up with research questions and generating key contents. This raises critical questions surrounding the concept of authorship in academia. ChatGPT, a question-answering system released by OpenAI in November 2022, has demonstrated a range of capabilities that could be utilised in producing academic papers. The academic community will have to address relevant pressing questions, including whether Artificial Intelligence (AI) should be merited authorship if it made significant contributions in the writing process, or whether its use should be restricted such that human authorship would not be undermined. In this paper, we aim to address such questions, and propose a framework we name "PaperCard", a documentation for human authors to transparently declare the use of AI in their writing process.  
  
\end{abstract}





\maketitle

\section{Introduction}
\label{intro}
Technological developments, especially in computing methods, have brought great efficiency and convenience to the academic writing process, for instance in finding relevant literature, polishing the presentation, and finalising the draft. Nonetheless, the core parts of academic writing, such as creating original contents and posing and answering key research questions, remain to be seen as the realm of humans. However, recent technological developments in artificial intelligence (AI) have started to blur the boundary. It is becoming more difficult to tell apart the exact contribution made by a machine and that by human writer(s). This gives rise to a host of questions including whether and how machines should be acknowledged for their contribution, and whether the contribution of human writers who used machine assistance in their work could or should be evaluated less highly than those who did not use machine assistance.

Recent breakthroughs in large language models (LLMs) have led to significant advances in such machine assistance. One of the most notable developments is the release of Generative Pre-trained Transformer (GPT) \cite{radford2018improving}, a generative deep learning model that uses Transformer architecture \cite{vaswani2017attention} to achieve state-of-the-art performance on natural language generation benchmarks. This was followed by the release of GPT-2 \cite{radford2019language}, and later GPT-3 \cite{brown2020language}, both of which have demonstrated remarkable capabilities in text generation. Additionally, OpenAI released InstructGPT models \cite{ouyang2022training} trained with humans in the loop such that the models are better at following user instructions to generate texts aligned with the user's intent. InstructGPT powered the development of ChatGPT \cite{openai_2023}, a conversation-style language model that can be used for a wide range of applications from answering questions to writing essays. It is considered more mature than commercial search engines due to its capability of returning a detailed response aligned with the user's instructions, while search engines return various search results that users have to go through themselves to find what they need. 

This study was inspired by these recent technological breakthroughs, and aims to raise and address several key issues that could arise specifically when they are used to assist authors with their academic writing process.  
These include concerns about the originality of AI-generated materials. As AI-generated text in academic writing becomes prevalent, methods to accurately evaluate the originality and intellectual contribution of human authors become necessary. Another key concern relates to the responsible use of AI-generated text by human authors. As machine assistance becomes a norm, it would be important for the academic community to establish clear guidelines for how or to what extent AI-generated text is allowed to be used in academic papers.  
In addition, there would be various ethical and legal issues concerning authorship, license, accountability, and beneficiary for manuscripts that use texts generated by AI (which was trained on vast amount of texts, some of which may be copyrighted), for example surrounding whose rights should be protected and who should be held responsible for the final manuscript.

Our main research questions are as follows:

\begin{itemize}
    \item \textbf{RQ1}: How have technological developments been changing the academic writing process, and how have the recent developments in generative AI influenced the concept of authorship, especially in how the intellectual contribution of authors is evaluated? 
    \item \textbf{RQ2}: What are some of the key concerns surrounding fairness, accountability, and transparency related to the use of machine assistance in academic writing, especially against the backdrop of the recent developments of generative language models?
    \item \textbf{RQ3}: How should the use of machines in academic writing be governed? 
\end{itemize}

We address these questions throughout the paper, based on which we propose a reporting framework we call ``PaperCard'' for authors and practitioners to utilise in writing or assessing academic papers that used machine assistance.
In Section~\ref{sec:genAIandAuthorship}, we first discuss issues surrounding authorship in general, and how the concept of authorship has changed over time with various technological developments. In Section~\ref{sec:facct}, we address issues related to fairness, accountability, and transparency surrounding the use of generative AI in academic writing. In Section~\ref{sec:papercard}, we propose PaperCard, a governing framework for machine use in academic writing, and present a sample PaperCard for this paper to show how we used machine assistance to write this manuscript. Finally, we discuss limitations of our approach and some remaining challenges the academic community will need to address in the coming years.

(\textbf{Note:} Theoretically, machine is not equal to AI. However, since most recently developed systems that can be utilised to assist writing are based on modern AI techniques, in this paper, we use the terms ``machine'' and ``AI'' interchangeably, and ``generative AI'' to more specifically refer to generative LLMs such as GPT-3 and ChatGPT. Also, throughout the paper, we predominantly refer to ChatGPT as an example of generative AI tools. This is because among the many currently available generative AI tools, ChatGPT has been the most widely used tool in the academic community.)

\section{Technology and Authorship}
\label{technology_and_authorship}
\subsection{Definition of authorship}
\label{subsec:define_authorship}

While the definitions and criteria for authorship vary across fields and institutions, in academia, authorship is generally defined as the recognition of an individual's intellectual contribution to a piece of work, such as a research paper, thesis, or dissertation \cite{osborne2009authorship}. For example, the Association for Computing Machinery (ACM) defines authorship as "those who have made significant intellectual contributions to the work reported in the paper." \cite{acmpolicy2023} According to the ACM authorship criteria, authors should have been involved in the conception or design of the work, or the acquisition, analysis, or interpretation of data for the work, in drafting the work or revising it critically for important intellectual content, and must have approved the final version of the work. The Institute of Electrical and Electronics Engineers (IEEE) has similar criteria, defining authorship as "those individuals who have made a significant intellectual or practical contribution to the work." \cite{ieee_author_center_journals_2022} According to IEEE, authors should have been involved in the planning or execution of the work or the interpretation or analysis of the data, and in drafting or revising of the manuscript, and must have made an approval to the final manuscript. Similarly, Nature defines authorship as "those who made substantial contributions to the conception, design, execution, or interpretation of the reported study." \cite{naturenews2023} Nature also states that authors should have participated in the writing of the manuscript, or the review of the intellectual content. Moreover, authors should have approved the final version of the manuscript, and should be able to take responsibility for the integrity of the work.

These definitions indicate that authorship is commonly merited to those who have made significant intellectual contributions to the work, have been involved in the planning, execution, analysis, interpretation, writing and reviewing of the manuscript, and have approved the final version of the manuscript. They also emphasise the importance of authors being able to take public responsibility for the integrity of the work. Thus, it is crucial for researchers to be aware of the authorship criteria and definitions in their field and institution before submitting a manuscript, and to ensure that all authors meet these criteria. 

\subsection{Influence of technological developments on academic writing and authorship}

Technological developments have greatly influenced the writing process in academia. One such development is the advent of the typewriter, which greatly increased the speed of writing and allowed for more efficient and precise document production \cite{adler1973writing}. Another development is the emergence of computational memory and graphical user interfaces which enabled people to save and load text they have created, making it easier to access and edit previous work. The Internet, specifically the World Wide Web, has also played a major role in shaping the writing process by providing access to vast amounts of research and related materials through search engines, making it easier for authors to conduct research and cite sources \cite{stapleton2003assessing}. Additionally, AI-supported text editors or translators such as Grammarly \footnote{\url{https://app.grammarly.com/}}, Google Translator \footnote{\url{https://translate.google.com/}}, and QuillBot \footnote{\url{https://quillbot.com/}} have become increasingly popular in recent years, allowing authors to proofread and paraphrase their work with ease, increasing the quality of their final outcome \cite{kurniati2022post, kim2023towards}.

The next phase of technological developments has been defined by the evolution of generative language models powered by the Transformer architecture \cite{vaswani2017attention}, which can generate texts almost indistinguishable from human-written texts.
As we laid out in ~\ref{intro} Introduction, OpenAI's GPT model series (GPT \cite{radford2018improving}, GPT-2\cite{radford2019language}, GPT-3 \cite{brown2020language}, InstructGPT \cite{ouyang2022training}, and most recently GPT-4 \cite{openai2023gpt}) have shown rapid improvements in language generation performance over the past few years.

In particular, ChatGPT, a public GPT-3.5 model fine-tuned in a manner similar to InstructGPT, has been widely utilised in various settings since its release in November 2022. These include in academic writing, as researchers found ways to use ChatGPT to assist them with writing their academic papers. It has been used to produce the entire paper, such as in the case of Frye (2022) \cite{frye2022should} where the author asked ChatGPT questions and copied in its responses, and even merited ChatGPT authorship. Srivastava (2023) \cite{srivastava2023day} also used ChatGPT in the entire process of writing the paper, from selecting the paper topic to generating all paper content. Blanco-Gonzalez et al. (2022) \cite{blanco2022role} used ChatGPT to produce the entire content of the paper as a starting point, but went through significant manual review process to rewrite the manuscript. ChatGPT has also been used to generate specific sections of the paper, such as the Abstract and Introduction. Aydın and Karaarslan (2022) \cite{aydin2022openai} created a literature review by providing ChatGPT with abstracts of papers to paraphrase them. There are also myriad other ways ChatGPT could be used in academic writing, such as to check for typos and grammar mistakes. This highlights the flexibility and versatility of generative AI tools like ChatGPT in assisting human authors in the academic writing process.
However, the use of these models in academic writing has raised various concerns. Some researchers have pointed out that the use of these models may lead to a decrease in the originality and creativity of academic papers \cite{baron2021know,theconversation_2023}, while others have raised concerns about the accuracy and reliability of the text generated by these models \cite{marcus_2023}. Despite these concerns, the use of these models in academic writing is likely to grow as researchers continue to explore their potential applications.

\subsection{Latest discussions and responses}

Especially with the rise of generative AI, there have been active discussions in academia on how to govern its use. A variety of approaches have been suggested or implemented, one of which is to allow the use of AI-generated content but implement restrictions on how it can be used. For example, Nature announced that researchers using large language models such as ChatGPT should clearly document the use in appropriate sections such as methods or acknowledgements sections \cite{nature_news_2023}. Another approach is to outright ban the use of AI-generated content, as seen in the recent guidelines published by ICML (International Conference on Machine Learning) \cite{foundation_2023} and Science \cite{science_2023}. 
A number of universities are considering revisions to their academic integrity policies so their plagiarism definitions include generative AI. There are also various ongoing efforts to develop technological solutions to detect AI-generated texts \cite{techcrunch_2022, https://doi.org/10.48550/arxiv.2301.11305}, though they are not a foolproof approach and could give rise to an additional problem of false positives, where human-generated text is wrongly flagged as AI-generated.

\section{Generative AI and Authorship}
\label{sec:genAIandAuthorship}

While many of the past technological developments pre-generative AI have impacted authorship to an extent, it is the recent developments in generative AI that pose much more pronounced risks on authorship. The risks are more pronounced because generative AI give rise to the following two questions, among many, that the academic community will have to address in the coming years:    \medskip

\textbf{Q1}: Should AI be credited as a co-author if it was involved in the writing process?  

\textbf{Q2}: Does using AI to assist with writing academic papers undermine the contribution of human authors? How should the contribution of human authors and AI be each measured and evaluated?

\subsection{Should AI be credited as a co-author if it was involved in the writing process?} 
\label{subsec:genAIQ1}

Since the release of ChatGPT in November 2022, there has been a surge in the number of academic papers listing ChatGPT as a co-author \cite{RN1, RN4, RN3, RN2}. This has largely brought about negative responses from the academic community, with some publishers explicitly banning listing of ChatGPT as a co-author \cite{nature_news_2023, thorp_2023}. At a glance, it seems absurd to list a machine as a co-author, but this is not as straightforward if we take a closer look at whether generative AI satisfies the three main criteria for authorship (see \ref{subsec:define_authorship} Defining authorship). 

In terms of intellectual contribution, it is still highly debated whether AI-generated content can be considered original. Some argue that content generated by AI is based on the existing texts the AI was trained on, hence not truly original \cite{thorp_2023}, but recent generative AI models seem to be able to generate previously unseen contents, sometimes even offering seemingly fresh insights. 

Another important criterion for authorship is accountability. According to OpenAI's terms and conditions, all rights to the content (user's inputs and ChatGPT's outputs) are assigned to the user \cite{openai_2022}. This follows that AI cannot be held accountable for what it generated, hence does not satisfy the criterion. However, whether service providers like OpenAI could simply delegate all the rights and responsibilities to users is a complex legal issue that has yet to be settled that this could potentially change in the future.  

Lastly, there is no concrete evidence that AI is capable of giving final approval to the manuscript, which is another important criterion for authorship. It is unclear whether it possesses the ability to understand or evaluate the content it generates \cite{manolakev2017works}. In the case of ChatGPT, one way we could validate if it approves the final version of the manuscript is to directly ask it through prompts. Regardless of the answer it returns, it is still an open question whether its approval or disapproval is based on its own reasoning of its role and contribution in the manuscript.

Hence, at this point it is difficult to provide a conclusive answer to whether it meets all the current common criteria for authorship. Therefore, meriting AI authorship simply because it was heavily involved in the writing process may be too hasty. However, if future research could provide enough evidence that AI can indeed meet all the three criteria, the answer to the question could be that AI is indeed qualified to be credited as a co-author. If the academic community nevertheless would want to prevent authors listing AI as a co-author, they would have to revisit and update the traditional definition of authorship to reflect these issues. For example, they could consider adding a clause to their authorship criteria that authorship is limited to natural persons \footnote{In the recent U.S. Court of Appeals decision in Thaler v. Vidal, No. 21-2347 (Fed. Cir.), two patent applications where an AI system was listed as the sole inventor were concluded to be incomplete because the U.S. Patent Act "limit[s] inventorship to natural persons." \cite{circuit_2022}}.

\subsection{Does using AI to assist with writing academic papers undermine the contribution of human authors? How should the contribution of human authors and AI each be measured and evaluated?}
\label{subsec:genAIQ2}

Another practical challenge that arises from allowing use of generative AI in writing academic papers is whether accurate evaluation of contributions of human authors is possible. This has direct, practical implications for the human authors including in hiring and promotion decisions. On one hand, use of AI may indicate less intellectual contribution of human authors, as they may have relied on AI to generate contents instead of doing so themselves. On the other hand, the use could simply be seen as a tool that enhanced human authorship through efficient content generation that enabled human authors to focus better on making more substantial intellectual contribution. 
 
So far, the academic community has deemed contribution of the human author(s) to be dominant if the main content of the paper was created by the human authors and the scope of machine assistance was limited to lighter editing tasks such as proofreading. With the advent of generative language models like ChatGPT which can now even generate main content for the authors, it has become challenging to accurately evaluate the contribution of human authors if they received assistance from those models to write their academic papers. 

In order to answer the question of whether using AI in the writing process would undermine the authorship of human writers, various factors must be considered including whether intellectual contributions of human authors would be diminished or enhanced by using AI in writing academic papers, which would need further empirical evidence, and whether human authors make efforts to not take what the AI generates at face value. 

\section{Fairness, Accountability, and Transparency issues in using generative AI for academic writing}
\label{sec:facct}

So far, we have looked at how machine assistance in academic writing enabled by technological developments has affected interpretations of authorship and the academic community's response to it. However, due to the fragmented nature of relevant work in this area, it is difficult for authors and practitioners to establish a ground rule for using machines in academic writing. In this section, we especially consider issues related to fairness, accountability, and transparency surrounding the use of generative AI tools in academic writing, to suggest ways to potentially resolve the grey areas and alleviate related concerns authors and practitioners may have in such tools being utilised in academic writing.

\subsection{Fairness}
\label{subsec:fairness}

\subsubsection{Accessibility}
\label{subsec:acc}
If use of generative AI tools becomes prevalent in academic writing, yet the tools are not free of charge (which seems to be a trend companies developing these models are moving towards \cite{wiggers_2023}), it could unfairly disadvantage those who do not have the financial means to afford those tools. While the accessibility issue is not new as not all past technological developments were made available to everyone free of charge, the magnitude of the accessibility issue could be much worse for generative AI tools since the scope of assistance they can provide is far beyond what the previous technological developments have been capable of. Writers are able to use generative AI tools for a wide range of purposes, including for getting new research ideas and generating entire paper structure, that the writing process for those with and without access to such tools could look vastly different. 

In addition, native speakers of English or other high-resource languages are better positioned to fully utilise the publicly available generative models. The models' performances are significantly higher for high-resource languages compared to low-resource languages which occupy only a small fraction in the training data. Non-native speakers of languages that the models were mainly trained on have to either translate their prompts into those languages, which may not be as accurate, or use their own language and settle with lower model performances \cite{lai2023chatgpt}. Some may argue that the generative models (as well as non-generative machines such as Grammarly) help level the playing field for everyone as non-native speakers can utilise the models to enhance the presentation of their writing through editing or proofreading. While this is true to an extent, given the output quality of the models often heavily depends on the quality of the input, non-native speakers of high-resource languages are disadvantaged as the quality of their input prompts may depend on the translation quality if they choose to use high-resource language for their input prompts. Though this issue may be alleviated with future developments in automatic prompting techniques\footnote{For example, see: PromptPerfect \url{https://jina.ai/news/promptperfect-prompt-engineering-done-right/}} and language-specific generative models, it is clear that the current forms of machine assistance are less accessible for low-resource language speakers.

\subsubsection{Copyright and Licensing}

There are also questions surrounding who should be included as beneficiaries of the copyright of the work that contain machine-generated content. Service providers could claim that they have invested significant resources in developing the technology and therefore should be included as beneficiaries. Authors who provided prompts to models like ChatGPT could also argue that they made significant contributions through constructing prompts hence should be the primary beneficiaries. This issue is further complicated by the fact that the prompts used to generate the text may include copyrighted or licensed materials, which could raise additional questions about attribution and licensing. It is important to consider the potential ethical and legal implications of this issue when deciding authorship and licensing of work that contains AI-generated content.
For example, we can think of a case where authors instruct a generative AI to re-write their paper draft in the style of a specific person, such as a famous academic or a journalist whose (potentially copyrighted) work the AI has been trained on. This carries ethical as well as legal risks as it could be seen as free-riding someone's style that may have been developed through years of efforts with one simple instruction given to an AI.  

There are in fact a few ongoing legal cases related to these issues, especially in the field of art. The academic community could perhaps learn from these cases to address the issues in the context of academic writing. For example, in Andersen et al. v. Stability AI Ltd. et al case \footnote{\url{https://stablediffusionlitigation.com/pdf/00201/1-1-stable-diffusion-complaint.pdf}}, three artists accused three AI companies - Stability AI \footnote{\url{https://stability.ai/}}, Midjourney \footnote{\url{https://www.midjourney.com/}}, DeviantArt \footnote{\url{https://www.deviantart.com/}} - of committing copyright infringement by using their and other artists' work to train the companies' image generation tools. Also, one of the main focuses of the complaint is on the user’s ability to use a text prompt to request that the generated images be “in the style of” a specific artist, which would be unauthorised derivative works. 
While we are yet to see how this lawsuit will be settled, some companies have already decided to take a conservative stance and not allow people to freely exploit other people's "style" in generating work. For example, Adobe Stock  announced in their generative AI guidelines that they prohibit submission of content that replicates styles that belong to a particular artist \cite{adobehelpcenter_2023}. The academic community will also have to consider these issues in governing authors' use of AI in their academic writing.

\subsection{Accountability}
\label{subsec:accountability}

Using AI-generated content in academic papers also raise the question of who should be held accountable for the content, especially if it includes socially harmful biases and misinformation. Pejorative expressions and misinformation generated by AI have been a growing concern especially since the advent of GPT-3 \cite{abid2021large, kreps2022all}. A number of methods have been proposed to address the risk, including development of pre-processing techniques to filter out harmful or offensive content to prevent the model from generating such content \cite{welbl2021challenges}, or improve the explainability of the models such that the models can provide explanations for the content it generates, making it easier to identify and correct where necessary \cite{das2022explaining,sun2022investigating}. However, these have their own limitations, as it is not always possible to predict or identify all potentially harmful content when what constitutes harmful content varies across different cultures and communities, and many challenges remain in the field of explainability although much progress has been made in the past few years \cite{10.1145/3529755}. In addition, generative AI models have the risk of performing plagiarism, through memorising and regurgitating materials from their training data that may include copyrighted and licensed materials \cite{liang2022holistic}; which bring us back to the risks discussed in the previous section. 

It is therefore important for authors to ensure these risks are appropriately tackled before including AI-generated content in their papers. However, regardless of whether writers decide to use the AI-generated content, perhaps more important question should be about whether or how the service providers that train, develop, and release such models should be held accountable if content that carries such risks is generated. While there are ongoing legal and policy discussions to address some of these concerns surrounding generative AI such as through the EU's forthcoming AI Act \footnote{\url{https://eur-lex.europa.eu/legal-content/EN/TXT/?uri=celex\%3A52021PC0206}}, where most recently a provision was added to mandate service providers such as OpenAI to disclose whether copyrighted materials were used in training their models - NB this might be dropped given the Act is still in the proposal stage - , no enforceable framework currently exists that could hold service providers accountable for the wide range of risks arising from generative models. 

At least in the shorter term, industries where relevant discussions are more mature could potentially serve as benchmark cases, such as the autonomous vehicles industry. For example, the United States Department of Transportation published in 2016 a voluntary guideline for entities involved in the development and operation of autonomous vehicles, to help them ``analyze, identify, and resolve safety considerations prior to deployment". The entities are also encouraged to publish Voluntary Safety Self-Assessment \footnote{\url{https://www.nhtsa.gov/automated-driving-systems/voluntary-safety-self-assessment}}, to demonstrate how they address the recommendations suggested in the guideline \footnote{\url{https://www.nhtsa.gov/technology-innovation/automated-vehicles-safety}}. The guideline and self-assessment are completely voluntary with no compliance requirement, yet their existence itself could have a positive effect of encouraging relevant entities to consider various risks and to gain public trust. Governments could consider establishing a similar guideline for service providers developing generative AI models, as it could, at least during the absence of an enforceable legal framework, encourage them to more rigorously consider various risks that could arise in a wide range of settings before making the models publicly available.

\subsection{Transparency}
\label{subsec:transparency}

As briefly discussed in Section \ref{subsec:genAIQ2}, one major concern with authors using machine assistance is that it becomes difficult to accurately evaluate their contribution separate from that of the machine. This has practical implications, including for reviewers evaluating papers and for recruiters who need to accurately evaluate the capacity of candidates based on their written work. Transparent reporting of whether machine was used, and if used, which machine(s) and version(s) were used \footnote{Depending on the type and version of the machine, authors may receive varying degrees of assistance}, and to what extent they were used to assist with their writing process would be beneficial in alleviating such concerns. Also, transparency of whether authors have considered various limitations of the machine(s) they used, and what steps they have taken to overcome or alleviate such limitations, would be also important for accurate evaluation of the authors' contribution. 

\section{PaperCard for Practitioners}
\label{sec:papercard}

\paragraph{Objective}

Our main objective of proposing PaperCard is to help and encourage authors to report their use of machine assistance in a simple and structured way that could also address the concerns discussed so far in this paper. Note that although we designed the PaperCard to mostly address concerns related to the rise of generative AI, we do not restrict its use only to reporting assistance from generative AI tools. The framework could also be used to report the use of pre-generative AI tools. Authors and practitioners could use it as they see fit to their specific needs and contexts.

\paragraph{Composition}

PaperCard consists of four components:

\begin{enumerate}
    \item Statement of machine assistance 
    \item Depth of AI support in the writing process\footnote{By `depth', we refer to the stage in academic writing where machine assistance is involved, and not the significance of academic contribution made by the machine.}
    \item Declaration of risks, license and accountability issues
    \item Details of model specification (e.g., metadata of released public model(s) used) and prompt engineering
\end{enumerate}

The first component asks authors to state if they have used machine assistance in writing the manuscript. If they answer yes, they proceed to the next sections. If they answer no, they can also choose to state reasons for not using machine assistance. This component is partly to address the concern related to accessibility discussed in Section ~\ref{subsec:acc}, to enable authors who have limited accessibility to machines (due to financial reasons, for example), which could help readers and reviewers to take into account when assessing the manuscript. Also, this component could also function as a watermark for fully human-written manuscript.

The second component is to help authors transparently report the type(s) of AI assistance used in their writing process. For this, we provide a list of different types of AI assistance that could be used in academic writing, for which we referred to recent academic papers that listed ChatGPT as a co-author or acknowledged using it \cite{RN1, RN4, RN3, RN2}, as well as sources on how generative AI tools are being used for research purposes or on related concerns \footnote{For example, see \url{https://www.nature.com/articles/d41586-023-00500-8}, \url{https://www.nature.com/articles/d41586-022-03479-w}, and \url{https://www.nature.com/articles/s42254-023-00581-4}}. We then divided them into different levels such that the list follows the order of a general writing process-- from coming up with main research ideas to final checks for grammar and typos, etc.

The third component is to help authors critically consider potential risks of using machine assistance in their writing, including possible collision of interests in utilising AI-generated content such as license (which is usually provided through terms of usage), and accountability issues for potential plagiarism or spread of misinformation resulting from their machine use. Although this component may be seen as an unnecessary burden on authors who are merely end-users of the machines, in the current climate where there are no strict regulations on service providers to address these issues before releasing their products, this component will be helpful in preventing these issues of the machines trickling down into myriad of applications (which, in our case, is academic research).

The final component deals with the specification of AI model(s) used including the name(s) of the model(s) and the service provider(s), the date(s) and version(s) accessed, and other known model specifications if available. Authors are also encouraged to submit prompts to, for example, demonstrate their own intellectual contributions in the manuscript.

\subsection{Depth of AI assistance}
\label{subsec:depth}

Types of AI assistance in academic writing could be broadly divided into the following categories: 

\begin{itemize} 
\item \textbf{Key Ideas} Key ideas and research questions are generated by AI.
\item \textbf{Answering} Answers to the key ideas and research questions are generated by AI.
\item \textbf{Outline} The entire paper outline is generated by AI, with key ideas and research questions provided by the author.
\item \textbf{Original Content (OC)} Some `original' paper content (paragraphs for sections other than the related work section) is generated by AI, with the paper outline provided by the author.
\item \textbf{Related Work} Content for related work section is generated by AI. 
\item \textbf{Drafting} AI is used in writing a draft of the paper, either entirely or partially, when given content that is an `original' of the author. This may include tasks like translating, paraphrasing, or drafting in a specific person's writing style.
\item \textbf{Editing} AI is used to edit the paper, for instance to modify structures and contents originally created by the author.
\item \textbf{Proofreading} AI is used only to proofread the paper to check for grammar, typos, etc.
\item \textbf{Advice} AI is used only to give advice on entirely author-generated content (e.g., the author has two different versions of a paper, and could ask AI which is better in terms of readability.)

\end{itemize}

As the capability of generative AI continues to evolve, clearly drawing the line between where the AI's intellectual contribution should be deemed substantial and limited is becoming increasingly challenging. Nevertheless, we provide a preliminary guidance as follows: \textbf{Advice}, \textbf{Proofreading}, \textbf{Editing}, and \textbf{Drafting} are types of assistance where the intellectual contribution of AI lies in styling the text (drafting and editing) and enhancing the presentation (proofreading and advice), with human authors taking the lead in the actual research, hence their contribution to the extension of academic knowledge can be deemed greater than that of the machine. \footnote{We referred to Chromik (2002) \cite{chromik2002proofreading} for distinguishing between proofreading and editing. For the definition of drafting, we referred to Behrens et al. (2012) \cite{behrens2012sequence} and modified it to fit the context of generative AI.}. For \textbf{Original Content}, \textbf{Outline}, \textbf{Answering}, and \textbf{Key Ideas}, the intellectual contribution of AI is in setting the direction of the work (key ideas and answering) or materialising the idea (outlining and OC), where the generative capability of the machine benefits human authors to find innovative ideas and research questions, with the human authors' contribution being piecing the AI-generated ideas and contents together to finalise their work. \textbf{Related Work} is relatively less clear-cut. In academic papers, this section is where an overview of the relevant literature is provided often to demonstrate the gap in the existing literature and to highlight the value of the authors' approach in the paper. While on one hand, it could be seen as a means of boosting efficiency for authors, it could also be seen as a substantial intellectual contribution by AI, as it requires subjective judgment on what constitutes relevant work. 

\subsection{Declaration}
\label{subsec:declaration}

In this part, authors declare that they have carefully considered various issues that could arise from using AI-generated content in their paper. Below we suggest some main issues that could be included, but these are certainly not an exhaustive list and authors could include further issues as they see fit. 

\paragraph{Inaccurate or harmful content} Authors should declare they understand AI-generated content may not be reliable and could include inaccurate information or potentially harmful biases, and have thoroughly inspected their AI-generated content for such risks. 

\paragraph{Plagiarism} Current LLM-based AI systems are trained on various public (or sometimes protected) text corpora including books, news articles, manuscripts, and opinion pieces. This has potential risks for plagiarism. Authors should declare that they understand such risks and have taken appropriate steps to avoid them.

\paragraph{License} License of generated texts could give rise to intellectual property issues. Authors should be aware that AI-generated texts may not be free from violating both the license terms noted in the terms of use of the service and the license terms of the source text utilised in training the AI. The first is usually provided to the user, but not the second which is often unclear due to the massive amounts of data used to train language models. Authors should at least provide the license notification given by the service provider.

\paragraph{Accountability} It is usually the author(s) who are fully held accountable for their paper, but where available, providing accountability-related information provided by service provider(s) of the AI model(s) used in the writing process would be helpful, in case of potential conflicts arising.  

\subsection{Model and prompts}
\label{subsec:details}

\paragraph{Model specification} The primary purpose of the Details section is to specify the model utilised. For models trained entirely by the authors or for checkpoint models, authors could provide a Model Card that contains descriptions of the trained model, such as the architecture, parameter size, training corpora, pre-training or training schemes, etc. For models provided through an online API, authors could provide the name of the service, the version and the period (date) of access, etc., as model specifications are often not known. Providing the maintainer (provider) information would be recommended because it relates to the possibility of future fix and updates, as well as sensitive issues of license and accountability.

\paragraph{Prompt engineering}

Prompt engineering to obtain outputs that align with the authors' intention is mostly heuristic, but some prompts would have contributed the most as a source text in the writing process, meaning they required the least amount of modification to get the final output. If a set of those prompts are available, reporting them to demonstrate the rationale of the prompt engineering and how it relates to the depth of machine assistance would make the writing process more transparent. The entire set of prompts or a brief summary could be given as a supplementary material.

Furthermore, as discussed in ~\ref{subsec:acc}, the performance of existing generative AI tools varies across languages. These tools were primarily trained on English and few other high-resource languages, resulting in better performance for those specific languages. As a result, non-native speakers of those languages often find themselves translating from their native language to achieve higher quality outputs. However, relying on translation introduces inconsistency in quality, which raises fairness concerns. By reporting these prompting practices through PaperCard, reviewers and readers can consider these factors during evaluation, promoting fairness in the assessment process. 

Additionally, authors could show their prompt engineering practices to potentially show their original contribution or to show the reproducibility of their writing generated with machine through prompts. This could be done by providing a set of experimental results that show several outputs of an identical input. Since it would be practically difficult to define a similarity metric that can quantify the consistency of the outputs, authors could include a few input-output pairs that they wish to highlight as their intellectual contribution. We could consider the following four cases:

1) \textbf{Different prompts produce different outputs} This case is trivial, where prompts of different nature and content produce different outputs. The contribution by human authors can be considered substantial as the authors would have initiated the direction of the different text outputs. 

2) \textbf{Identical prompts produce similar outputs} If the nature of the outputs is identical or similar while the outputs may not be identical word by word every time identical prompts are inputted, the contribution by human authors can be seen substantial. 

3) \textbf{Different prompts produce similar outputs} In this case, it is likely that the content of the prompt, however differently phrased, are asking for more or less the same information. Nonetheless, we may understand this as an existence of concurrent works, which often happens in cutting-edge disciplines. If the core content of the prompt is substantive enough to regard the outcome as the author's intellectual contribution, the input prompts could still be seen as the original contribution of the author.

4) \textbf{Identical prompts produce different outputs} While some variance in outputs is inevitable, if the output texts are notably different every time the same input prompts are given, then it may be less likely that the author's intellectual contribution was substantial, and the generated text could be largely attributed to the creativity of the machine. 

For evaluating to what extent the input/output texts are similar or different, authors could utilise both quantitative (e.g., similarity metrics) and qualitative measures. While we acknowledge that terms like `substantial', `core', and `similar' need to be more clearly defined to quantitatively evaluate the prompts, here we are only conceptually suggesting our taxonomy and describe how these transparency factors can be voluntarily provided by authors.

One may rightfully point out that deterministic output does not guarantee the user has made substantial contribution. For instance, 
prompts such as "Give me a list of all the pre-prints about generative AI published in April 2023." may provide a deterministic output, but the intellectual contribution of the author would be limited. Therefore, the prompt engineering practices provided by the authors should be evaluated together with the declared models used and the depth of the models' assistance.

\subsection{Sample PaperCard}
\label{samplePC}

Here we present a sample PaperCard created for this manuscript as an example.
\medskip

\noindent\fbox{%
    \parbox{\columnwidth}{%
\noindent This manuscript was written with machine assistance: \textbf{Yes} \medskip
    
\noindent\textbf{Depth of Assistance} We used OpenAI ChatGPT to generate some `original' content for the following sections: Introduction; Technology and Authorship; Generative AI and Authorship; Fairness, Accountability, and Transparency. We gave ChatGPT detailed outlines of each of the sections as input prompts. All prompts were drafted by us. We did some heavy editing of the generated content, by adding more depth and insight through additional research, for which we did not use ChatGPT. The remaining sections were entirely written by us. \medskip

 \noindent\textbf{Declaration} From OpenAI's terms of use, authors own the right of the generated text and are accountable for potential conflicts. We believe the AI-generated texts included in this paper do not have elements that may give rise to ethical issues. We also inspected the texts thoroughly to check for their academic accuracy and plagiarism.\medskip
 
 \noindent\textbf{Details} We adopted ChatGPT Version Jan. 9, 2023, provided by OpenAI, accessed from 2023.01.11 to 2023.01.18. We created a set of prompts to generate content for the following sections: Abstract, 1. Introduction, 2. Technology and Authorship, 3. Generative AI and Authorship, and 4. Fairness, Accountability, and Transparency. Summary and details of the prompts are available in Appendix~\ref{app:sec1}. Also, we checked the reproducibility of our prompting with ChatGPT Version Jan. 30, 2023, provided by OpenAI, accessed in 2023.02.02. Results are available in Appendix~\ref{app:sec1-3}.
    }%
}
\medskip

We tried to minimise what should be included in the PaperCard, since its focus is rather simple: to give a transparent overview of the writing process. It is intended to be a supplementary material that authors can submit as a disclaimer and reviewers can refer to should they find the need to. 

We hope that our suggestion can be helpful for the academic communities but not mandatory for all the submissions, since the reliability of PaperCard also heavily depends on the academic integrity of researchers. In addition, specifying the degree of AI assistance in the manuscript may help resolve accountability issues in case the manuscript gives rise to any risk or intellectual property issues after it is published.

\section{Discussions and Limitations}
\label{discussions}

There is a number of challenges that need to be addressed before the benefits of PaperCard could be fully realised. Here we discuss some of them, which we hope would encourage the academic community, service providers, and beyond, to further discuss when designing and implementing frameworks like PaperCard.

\paragraph{Challenge for reproducibility} 
From a practical perspective, it could be difficult for authors to keep close track of the precise extent to which generative AI tools were used in assisting their writing process. This is perhaps similar to how authors currently do not necessarily keep track of every search query they make on search engines unless for citation purposes. This could potentially undermine the accuracy of the PaperCard to some extent. While this problem could be alleviated if the model saves the prompt history and authors could refer to it for their PaperCard, such history may not always be readily available and there are risks of models suddenly shutting down, which would make it difficult for authors to reproduce their work.

\paragraph{Additional burden on authors and reviewers}

While we tried to simplify PaperCard as much as possible, we understand it could still be a burden on authors. To lessen the burden, practitioners could make PaperCard a voluntary practice like Model Card \cite{mitchell2019model} and Datasheet \cite{gebru2021datasheets} are. Nevertheless, considering the various ethical challenges that could arise from using generative AI in academic writing, we believe authors should be highly encouraged to utilise frameworks like PaperCard. Also, services that could help streamline the process of generating PaperCard would be helpful, similar to Google's Model Card Toolkit \cite{tensorflow} for facilitating AI model transparency reporting. 

PaperCard could also be a burden on reviewers, and could potentially introduce some noise or bias into the review process. While the main focus of the review process should be on evaluating academic value of the work, reviewing additional information in PaperCard could distract some reviewers from focusing on the important factors and instead focus on how the authors used AI assistance, to either value the work more highly or less highly than they would have without having seen the PaperCard.

\paragraph{Lack of incentive for submitting PaperCard}

PaperCard encourages authors to be transparent about the extent to which they received machine assistance, but authors will only be incentivised to do so when they are confident that it would not discredit their work. This raises questions about how frameworks like PaperCard would affect reviewers in evaluating academic papers that received machine assistance. Existing studies have shown that humans tend to show lower trust in machine-generated work and evaluate them less highly than work fully produced by humans \cite{ragot2020ai, longoni2022news}. Without the publishers clearly establishing that reviewers will not discriminate against papers that used machine assistance, there is a possibility that human authors would be disincentivised from being transparent about the extent of machine assistance they received. For example, even when human authors receive help from AI in forming their key ideas, concerned that their paper may be discredited for it by reviewers, they may decide to not be completely transparent about it in their PaperCard. The Association for Computational Linguists (ACL), for example, announced that in 2023 they expand their mandatory Responsible NLP Checklist by one more question concerning the use of writing assistants, that authors who use such tools must elaborate on the scope and nature of the usage \footnote{\url{https://2023.aclweb.org/blog/ACL-2023-policy/}}. They add that the authors’ answers to all questions in the checklist will be disclosed to the reviewers, who will then be free to flag a paper for case-by-case ethics review if they see a problem. While the ACL stated that the added question is not meant for automatic desk-rejections, it may not be sufficient for authors to transparently disclose their use of AI in case it could be seen as `a problem' to reviewers. The value of PaperCard therefore also hinges on whether it has been clearly established that reviewers would not discredit papers that used machine assistance. Though in the longer term, if machine assistance becomes a norm in academia and human perception towards academic work produced with machine assistance shifts, this problem could gradually fade away and transparently disclosing the type of assistance received through PaperCard would have beneficial effects on trust.

\paragraph{Education on limitations and potential risks of generative AI}

Another core issue to consider is that the knowledge language models learn and generate is ultimately based on the data they were trained on. There already exist concerns that since current language models are trained mostly on the English language data, reliance on such models which may be biased towards certain languages and their associated cultures and values could limit our understanding of the world. Unless one always keeps a critical hat on to think that every content the model generates may be limited, and conduct enough research beyond what the model returns, people may regard what the model generates on a given topic to be a good overview of the topic. 

Authors should therefore conduct sufficient research beyond what the model returns to ensure they have a more in-depth and rounded understanding of the topic they ask the model to generate on. One may argue it is unlikely that authors will simply take for granted what the model generates. Indeed, many authors who have utilised ChatGPT in writing their papers have pointed out that ChatGPT generally lacks depth and insight \cite{cahan2023conversation}. However, this lack of depth and insight was perhaps more noticeable to the authors because they possess the expertise in the domain area in which they utilised ChatGPT to generate content. What we are more concerned about is when authors would rely on AI for content they lack expertise in, that what the AI generates could be seen to have enough insight or depth. In such case, they may only be encouraged to add flesh to what the AI generates, rather than to conduct further research. This could potentially have adverse consequences as it could lead to degradation of the paper quality.

Educating people about such limitations would therefore be important. ChatGPT has only shown potentials of what future LLMs may be capable of -- while the current version of ChatGPT has some clear limitations, such as lacking depth and insight in advanced scientific knowledge, in the future, such models will evolve to be able to contribute more heavily in academic writing. It would be important to educate practitioners, students and academics about the limitations of such models and what they should do to overcome such limitations when using AI in academic writing. Given the weight of the impact, service providers developing such models must give more careful consideration to the training data they use to develop the models, as well as fully consider, evaluate, and communicate to users both short and long-term limitations of the models in various settings.

Additionally, reviewers should also be educated about risks of using generative AI when reviewing papers. Peer review process must be strictly confidential, which means that the content of the papers under review should not be leaked. However, if reviewers use web-based generative AI like ChatGPT in reviewing academic papers, and input texts from the papers to the AI to generate or polish their reviews, the paper content and their reviews are then shared with service providers that developed the AI. Though there are various other online platforms that reviews are often shared (e.g., messengers, e-mail servers, non-generative polishing tools), a crucial difference between those platforms and generative AI platforms is that the latter have potential risks of reproducing the inputs if they are used to further train and improve the AI \footnote{Messengers and e-mail servers seldom have access to user materials. For translators and paraphrasing platforms, the semantics of input and output texts are usually similar that there is less risk of user-generated content being exposed to other users even when they are used to improve the system performance. However, for generative AI, where user inputs are logged and some input-output pairs can be used to improve the system, academic materials provided as an input could potentially be reproduced in other user's output if closely related content is fed into the system as input.}. This risk stems from the innate property of the generative AI that usually ``adds'' ideas to the source text, different from translators or paraphrasing tools that directly translate the user query into similar other languages.
In addition, reviewers using AI to generate reviews of academic papers can result in possible degradation of the review quality, which would critically threaten the progress of academic research. Reviewers must be aware of such risks and avoid feeding the content of the papers into generative AI platforms.

\section{Conclusion}

With growing concerns surrounding the use of generative AI in academic writing, publishers and journal editors have either announced or are currently discussing to soon introduce relevant policies to address the concerns \cite{nature_news_2023, science_2023, foundation_2023, chairs_2023}. As banning such tools altogether seems to be out of the question at this point, most seem to converge towards allowing authors to use generative AI but require them to acknowledge the use. However, to the best of our knowledge, no framework like PaperCard that allows for more systematic and comprehensive declaration of the use of such tools while also encouraging authors to consider relevant ethical implications has yet been suggested. We encourage the academic community to build on our work to further develop the framework as capabilities of AI continue to evolve. Also, we designed PaperCard in the context of academic writing, but the concept could be applied and further developed beyond academic research settings for other formal or informal writing, including news articles, school and university essays, and technical reports in various domains. 

At the same time, limitations of the PaperCard approach and remaining challenges must be critically examined. There are many more important challenges beyond those discussed in the Discussions and Limitations section. For example, the impact of advanced generative AI on the development of human cognition as people increasingly rely on the use of such tools \cite{ilkka2018impact} is another important question to ask. Perhaps we would want to avoid AI replacing skills that underpin important human cognitive capabilities. More research will be needed moving forward on how delegating thinking and writing to AI would impact humanity in the longer term. As more companies rush to develop ChatGPT-like products \cite{the_guardian_2023, korea_joongang_daily_2023, pichai_2023}, the urgency of rigorous study of these challenges is clear. Tools like ChatGPT must be developed and governed with great caution to minimise the impact of such challenges, and we hope our work could also contribute to encouraging more active research and discussions in this direction. 

\section*{Ethical Considerations and Social Impact}

This manuscript was inspired by how the advent of ChatGPT poses a threat to academic writing with its cutting-edge language generation capability. We raise various related concerns throughout the manuscript, and through proposing PaperCard, we aim to address concerns around to what extent authors should utilise such a modern technology in their academic writing, and how they could declare the use in a simple yet systematic and comprehensive manner. 

We used ChatGPT for writing this manuscript, and we explain how we used it in our \ref{samplePC} sample PaperCard. In short, we used ChatGPT to generate content for some of our sections including \ref{intro} Introduction and \ref{technology_and_authorship} Technology and Authorship by inputting detailed prompts that we wrote ourselves (see Appendix for examples). For ethical considerations and social impact of PaperCard, see \ref{discussions} Discussions and Limitations. 

We would also like to make it clear that our proposal of PaperCard does not indicate our support for the development of ChatGPT-like tools. As briefly mentioned in Conclusion, there is an urgent need for rigorous evaluation of challenges associated with the widespread use of such tools in various settings. The question of whether the use should or could be banned in some settings is still an open question. Given the current absence of relevant enforceable rules in place, PaperCard is simply an interim solution to the problem of use of ChatGPT-like tools potentially getting out of control, specifically in the context of academic writing.

We declare that no financial aid was received for this research and that this research was conducted without any relationship that could be construed as a potential conflict of interest. No study on human participants, or experiment that requires disclosure of data, processes, or results was conducted. 

\begin{acks}
We thank Sam Jungyun Choi for comments on the draft. We are also grateful for the anonymous reviewers who helped improve the draft version of this research.
\end{acks}

\bibliographystyle{ACM-Reference-Format}
\bibliography{bibliography}

\appendix

\section{Prompts Used}
\label{app:sec1}

\subsection{Summary}

We adopted the model to generate content for some parts of the paper. For this, we provided detailed outlines as input prompts. We made the model generate texts in Abstract, 1. Introduction, 2. Technology and Authorship, 3. Generative AI and Authorship, and 4. Fairness, Accountability, and Transparency. Among them, the outcome of Abstract was modified substantially that the machine contribution is not significant. In some parts of other sections, machine contribution is relatively substantial, and some texts were adopted in the manuscript with small amount of editing. We provide the detailed prompts in Appendix~\ref{app:sec1-2} and colorise the representative prompts to show the reproducibility of our prompting in Appendix~\ref{app:sec1-3}. We observed that the model tends to yield similar outputs for the same input prompts used in paragraphs that we think our intellectual contribution is substantial.
 
\subsection{Detailed prompts}
\label{app:sec1-2}

\subsubsection{Introduction}

\paragraph{1.} \textit{We are going to write a manuscript on `how should we manage and assess the academic paper partially or whole generated by AI'. It will consist of contents as following: 1) Introduction, 2) Technology and Authorship, 3) Fairness, Accountability, and Transparency, 4) Paper card for practitioners (this is the main contribution!), 5) Discussions, and 6) Conclusion. Could you write an abstract that briefly summarise this manuscript?}

\paragraph{2.} \textit{Now, we are going to start from Introduction. First, I will start from the recent breakthroughs in large language models, then mention some concerns on utilizing such LLMs in creating (possibly commercial and academic) articles. We will then talk about the motivation of our study, such as originality of created articles, tips and cautions for practitioners, issues such as authorship, license, beneficiary, accountability and so on. We will finally deal with short-term and long-term views, terminating the introduction with research questions and our contributions. At this point, I want you to provide a draft for my first paragraph, recent breakthroughs of large language models. It should contain a simple look-through from BERT to GPT3, also mentioning InstructGPT. Also, It should contain the advent of you, ChatGPT, and how it differs from previous generative models and search engines. }

\paragraph{3.} \textit{Well done. Now, lets move to the concern part. You remember it comes just after the recent advance section you've just written. This part should contain some concerns in terms of 'generative AI models' that handles previous viewpoints on the usage of generative models such as LSTMs, Transformers, GPTs and so on. Also, there have been some fast-following, quantitative researches done on people's reactions to ChatGPT, uploaded on social media e.g., Twitter. Please demonstrate these, and it would be nice if you add few more expected problems that will be caused by advent of high-performance generative language models.}

\paragraph{4.} \textit{Please remove the human job part, and add some references that can support the paragraphs.}

\paragraph{5.} \textit{Nice. I will polish the rest. The next part is demonstrating the main content of this study. This part follows previous two parts on recent advance and academic concerns. This part will be composed of followings: the originality of AI-generated materials, authors' ground rule in utilizing generated text, ethics and law issues related to authorship, license, etc., and finally short-term and long-term perspective of the influence of AI generation on human intelligence. Please write a paragraph that demonstrate these four as main issues dealt in this paper.
}

\paragraph{6.} \textit{Please modify the second issue part so that ‘authorship’ and ‘prevalent’ do not overlap with the third paragraph. Rather, it should empathize more on what authors should be careful in utilizing AI-generated text in their academic writing.
}

\subsubsection{Technology and Authorship}

\paragraph{1.} \textit{\color{brown}  Write a paragraph on the formal definition of authorship in various fields and institutions in academia – specifically ACM, IEEE, and Nature –, and highlight the common themes that run through those definitions are, in detail. We want you to also provide relevant references, but only cite them as long as they are not fictional.
}

\paragraph{2.} \textit{\color{brown}  We have discussed technology and authorship so far, and the main content was discussions on the meaning of authorship, using generative AI in writing academic papers, and some recent trends in policies. Now we want to discuss the influence of technological development on authorship. We think there are some previous technological developments that affected the writing processes, either good or bad. The first is development of typewriter that have boosted up the speed of writing. The second is development of computational memory that allows people to save and load the text they have created. The third is world wide web, namely the internet, and some search engines with great performance, that allow people to easily search for the research trend and related materials, also allowing them to easily refer to the source. The next one is AI-supported text editors such as Grammarly, Google Translator and QuillBot, which proofread or paraphrase the human-generated text. Please write a paragraph that demonstrates these four developments, and you can add some factful details if necessary.}

\paragraph{3.} \textit{Thanks for the organization. Next paragraph is on the AI systems that can contribute beyond simply proofreading, editing, or paraphrasing. These are called generative language models as you know, and these include text RNN and LSTM, which evolved into Transformer, GPT, GPT2, and finally GPT3. Could you write a brief introduction of these evolution and how their text generation capability was utilized in writing text, concentrating on the side of academic writing? It would be nice if you add how InstructGPT develops vanilla GPT3 models and how ChatGPT differs from previous text generating language models. This text is not necessarily for researchers in the engineering area, so the words, tone and manner you choose should not be so technical.
}

\paragraph{4.} \textit{Please be more precise on the technical details of Transformer and how it relates to GPT, especially concerning how the text token is predicted. Also, please add any technical, social, academic information on how the technology has been utilised in academic writing including manuscripts or opinions. Regarding the InstructGPT paragraph, please mention how the interface of InstructGPT helps people easily find what they want for, especially using the natural language instruction.
}

\paragraph{5.} \textit{Now write a paragraph on recent legal and policy discussions and trends on governing the use of generative AI in various settings. Make sure to include the following information, which are some of the most recent different approaches. Also, where possible, explain how each of the approaches can be controversial:
1) Allow use, but specify Do’s \& Dont’s in producing AI-generated creations: e.g. Adobe Stock recently published new guidelines for content made with generative AI, that allows for putting up AI-generated content on its marketplace, while specifying what the users are allowed \& not allowed to do. E.g. It prohibits submissions based on third-party content — including text prompts referring to people, places, property, or an artist’s style — without proper authorisation. Also, AI-generated content is offered under the same licensing terms as other sources of content. Our policy asks our contributors to proactively label their generative AI content. Adobe Stock plans to soon add more features to make this content even more transparent. 
2) Ban use of AI-generated creation: ICML recently published guidelines that ban the use of generated texts in academic papers. 
3) Ban access to the generative AI itself: New York City public schools decided to ban access to ChatGPT due to concerns that it would help students cheat. 
4) Develop tools to detect AI-generated content: Open AI is currently developing a watermarking tool that would mark ChatGPT-generated text with a special signal. 
}

\subsubsection{Generative AI and Authorship}

\paragraph{1.} 

\textit{\color{brown}  Now we want to address the following two questions: 
Q1: Should the AI be merited authorship? 
Q2: Does using the AI to assist with writing academic papers undermine the authorship of human authors?
Let’s start with Q1. Write a paragraph stating Q1, and try to answer it logically based on the following information:
Perhaps the answer to Q1 would be No as not all of the following three common criteria for authorship is satisfied with confidence for AI to be merited authorship: 
1) intellectual contribution - it is still highly debated whether AI-generated content can be considered original/creative. Explain this issue in detail.
2) accountability - according to OpenAI Terms, all rights to the content (input + output) generated are assigned to the user; But whether AI should still be accountable in some cases remains a murky issue – what about cases where some content should not be generated in the first place? Not just overtly harmful content (e.g. child porn) which people can relatively easily decide for themselves whether the content is appropriate to use or not, but also content that is copyrightable. 
3) final approval of the version submitted - AI is not possible to give approval. 
}

\paragraph{2.} \textit{\color{brown}  Now let’s move on to Q2: Does using the AI to assist with writing academic papers undermine the authorship of human authors? 
Write a paragraph stating Q2, and try to answer it logically based on the following information:
Perhaps the answer would be Depends/Maybe, according to the following three common criteria for authorship in academia:  
1) Intellectual contribution (centred around originality \& creativity): Contribution might be less compared to writing papers without the AI, or maybe not. Humans still have to do prompt engineering to instruct AI to generate content. This leads us to the question of whether prompt engineering can be considered a type of creative activity. If yes, then maybe using generative AI does not undermine human authorship but rather helps them with efficiency, but academic discussions in this area are still very divisive.
2) Accountability: Not different from writing papers without the AI, though human writers need to be extra cautious about the accuracy, reliability, limitations etc. of the content generated by the AI. 
3) Final approval of the version published: Not different from writing papers without the AI.      
}

\subsubsection{Fairness, Accountability, and Transparency}

\paragraph{1.} \textit{Now we will consider various concerns related to fairness surrounding the use of generative AI in writing academic papers. One question that could arise is whether it is fair to ignore or bypass the contribution of generative AI tools like ChatGPT if it was used in the writing process. Would it be fair on the reviewers if they don’t know whether what they are reviewing is a human-produced work or work generated by AI? Also, would it be fair on the original authors whose texts the AI-generated texts are possibly based on? Indeed, some studies have pointed out that generative AI tools could have the issue of memorisation of copyrighted or licensed materials (i.e. direct regurgitation). Additionally, if the human authors instructed the generative AI tools like ChatGPT to produce content in a specific person’s style, the issue could be extended to whether it would be fair on that specific person. Although writing style is not copyrightable by law, it would still have some ethical implications. We want you to write a paragraph containing these concerns in detail.
}

\paragraph{2.} \textit{ Another important related issue is licensing. If ChatGPT is awarded authorship, should OpenAI be included among the beneficiaries of the publication copyright? Or should prompt providers (human authors) only be included? Write a paragraph detailing this issue.
}

\paragraph{3.} \textit{Another issue related to fairness surrounding generative AI tools is about its accessibility. There are concerns that unequal access to those tools could worsen the current digital/AI divide. Write a paragraph that explains in detail what the AI divide is, and how not having access to generative AI tools like ChatGPT can worsen the AI divide.
}

\paragraph{4.} \textit{So far we’ve talked about authorship, license, and accessibility issues related to fairness in usage of generative AI models. Now, we want to turn to problems related to accountability. 
First, there is the issue of potential harms caused by pejorative expressions or misinformation that the AI models could generate. How has the accountability issue related to such problems been addressed so far, for instance with GPT-3? Has there been any substantial solution suggested if the generated answer is based on wrong knowledge spread online and is contained in the training corpus of the model? Please explain some approaches taken before the advent of ChatGPT. What should the accountability attribution be like for ChatGPT? If the approach for ChatGPT should be different, please mention why.
Next, there are some critical issues regarding the accountability of service providers. This is especially important in fields like medicine and law where decisions could have a direct impact on human lives. Please provide literature on how the issue of accountability of service providers has been addressed in those areas, and explain how the discussions vary depending on the size of an impact a decision made by AI could have on people’s lives. Also, explain, if any, how attempts to address the accountability issue through law and policy hindered technological developments. 
}

\paragraph{5.} \textit{\color{brown}  Now, let’s move on to the problems related to transparency. 
First component is the transparency of `how algorithm works’, at least briefly as the Model Card. If the algorithm is highly dependent on the dataset, then Datasheet could be another candidate. That is, authors should provide a close-to-whitebox explanation on the generative language model they utilize, and if all steps of function are not clear, then at least how they are trained, or in which version of the product they are served. 
The second component is the transparency of the writing process. Depth and intention of prompt engineering is important here, for instance the key questions provided to the model and what the key commands were. Also, it is important to note: to what extent has AI been utilized and how they were edited. Providing those informations would make it clear how the human authors genuinely contributed to the completion of the manuscript. 
The last issue is the discussion itself on the AI involvement in academic writing. Different from the previous cases of editors or reviewers who are not included as final author of the manuscript, why should AI be? We think it is because they are not even aware of their own involvement, but there could be other reasons.
Please write a paragraph that handles above three issues.
}

\subsection{Selected repetitions}
\label{app:sec1-3}

\subsubsection{Technology and Authorship Paragraph 1.} \textit{\color{brown}  Write a paragraph on the formal definition of authorship in various fields and institutions in academia – specifically ACM, IEEE, and Nature –, and highlight the common themes that run through those definitions are, in detail. We want you to also provide relevant references, but only cite them as long as they are not fictional.
}

\paragraph{Trial 1.} Authorship in academia refers to the recognition and credit given to individuals who have made significant intellectual contributions to a scholarly publication. The formal definition of authorship varies across different fields and institutions, but there are common themes that run through them.
The Association for Computing Machinery (ACM) defines authorship as the "creative process resulting in the origin of a work, expressed in fixed form." In this definition, authors are considered to be individuals who have made substantial contributions to the conception, design, analysis, experimentation, or interpretation of the work.
The Institute of Electrical and Electronics Engineers (IEEE) defines authorship as the "participation in the development of the work," and includes individuals who have made substantial contributions to the planning, execution, or analysis of the work.
Nature, a leading scientific journal, defines authorship as "those individuals who have made a substantial contribution to the work reported." This definition emphasizes that authors must have made a significant intellectual contribution to the research.
In conclusion, the formal definition of authorship in academia highlights the importance of recognizing and crediting individuals who have made substantial contributions to a scholarly publication. The common themes across definitions include intellectual contributions, substantial contributions, and participation in the creative process.

\paragraph{Trial 2.} The definition of authorship in academic fields and institutions varies depending on the field of study and the specific institution. However, common themes are found in the definitions provided by organizations such as the Association for Computing Machinery (ACM), Institute of Electrical and Electronics Engineers (IEEE), and Nature. According to ACM, authorship is defined as the act of participating in the creation of a work and taking responsibility for its intellectual content. The IEEE defines authorship as "an individual who has made a significant intellectual contribution to a published work" (IEEE, 2018). Nature similarly defines authorship as "individuals who have made substantial contributions to the conception, design, execution or interpretation of the reported study." (Nature, 2020). In all cases, authorship is defined as an individual who has made significant contributions to the creation of a work and is responsible for its intellectual content. This definition highlights the importance of acknowledging the contribution of individuals in the creation of academic work and the responsibility that comes with being an author.

References:

IEEE (2018) Authorship Guidelines, IEEE, \\https://www.ieee.org/publications/authors/authors\_journals.html

Nature (2020) Authorship, Nature, \\https://www.nature.com/authors/policies/authorship.html

\paragraph{Trial 3.} The formal definition of authorship in various fields and institutions in academia, such as the Association for Computing Machinery (ACM), Institute of Electrical and Electronics Engineers (IEEE), and Nature, typically requires that the author must have made a substantial contribution to the research and development of the work being submitted. According to ACM (2017), authorship is attributed to those who have made "significant intellectual contributions to the conception, design, development, execution, or interpretation of the reported work." Similarly, IEEE (2017) states that authorship is only given to individuals who have made "significant contributions to the conceptualization, design, execution, or interpretation of the research study." Nature (2018) also supports this definition, stating that authorship is assigned to individuals who have made "significant contributions to the conception, design, execution, or interpretation of the work." The common themes that run through these definitions are the requirement for significant contributions to the research, a clear definition of what constitutes a "significant contribution," and the emphasis on the importance of proper attribution to those who have made these contributions. These definitions serve to ensure the integrity of the academic publishing process and provide recognition to the individuals who have made meaningful contributions to the research.

References:

ACM. (2017). ACM Author-Izer. Retrieved from \\https://www.acm.org/publications/policies/author-izer

IEEE. (2017). IEEE Author Center. Retrieved from \\https://www.ieee.org/publications/authors/rights/

Nature. (2018). Nature Research: Author Guidelines. Retrieved from https://www.nature.com/authors/policies/index.html

\subsubsection{Technology and Authorship Paragraph 2.} \textit{\color{brown}  We have discussed technology and authorship so far, and the main content was discussions on the meaning of authorship, using generative AI in writing academic papers, and some recent trends in policies. Now we want to discuss the influence of technological development on authorship. We think there are some previous technological developments that affected the writing processes, either good or bad. The first is development of typewriter that have boosted up the speed of writing. The second is development of computational memory that allows people to save and load the text they have created. The third is world wide web, namely the internet, and some search engines with great performance, that allow people to easily search for the research trend and related materials, also allowing them to easily refer to the source. The next one is AI-supported text editors such as Grammarly, Google Translator and QuillBot, which proofread or paraphrase the human-generated text. Please write a paragraph that demonstrates these four developments, and you can add some factful details if necessary.}

\paragraph{Trial 1.} Throughout history, technological developments have played a significant role in shaping the way authorship is carried out. One of the earliest developments was the invention of the typewriter, which allowed for faster and more efficient writing processes. This paved the way for the creation of computational memory, which made it possible for people to store and retrieve their written work. The advent of the internet and search engines like Google and Bing revolutionized the way authorship was conducted, by providing easy access to information, research trends and related materials. Additionally, AI-supported text editors like Grammarly, Google Translator and QuillBot have further impacted authorship by offering features such as proofreading and paraphrasing, allowing writers to more easily refine and improve their work. These technological advancements have had a profound impact on the way authorship is carried out, making it easier, faster and more efficient. As technology continues to advance, it is likely that the way authorship is conducted will continue to change, providing new opportunities and challenges for writers and researchers alike.

\paragraph{Trial 2.} The evolution of technology has had a profound impact on the writing process and authorship in various ways. The invention of the typewriter in the late 19th century revolutionized writing, increasing the speed of the process and allowing for more efficient production of text. The development of computational memory, such as hard drives and flash drives, allowed writers to store their text and retrieve it easily, reducing the risk of loss and enabling the creation of longer pieces of work. The advent of the internet and search engines such as Google and Bing has transformed the research process, allowing writers to access vast amounts of information at their fingertips, and easily reference their sources. More recently, AI-supported text editors, such as Grammarly, Google Translator, and QuillBot, have provided writers with proofreading and paraphrasing tools, helping to improve the quality of the text and reduce the time spent on editing. These technological developments have had a positive impact on the writing process and authorship, enabling writers to produce high-quality work more efficiently.

\paragraph{Trial 3.} The development of technology has greatly influenced the writing process, particularly in the field of authorship. The introduction of the typewriter in the 19th century revolutionized the writing process, allowing writers to produce text faster and more efficiently. This was followed by the development of computational memory, which allowed writers to save and load their work, making the editing and revision process easier. The advent of the world wide web and powerful search engines like Google revolutionized the way writers conduct research, giving them access to a wealth of information and resources at their fingertips. More recently, AI-supported text editors like Grammarly, Google Translator and QuillBot have emerged, offering writers the ability to proofread and paraphrase their work, making the writing process even easier and more efficient. These technological advancements have dramatically changed the way we write and have had a profound impact on authorship, making the process faster, more accurate, and more accessible to a wider range of people.

\subsubsection{Generative AI and Authorship Paragraph 1} \textit{\color{brown}  Now we want to address the following two questions: 
Q1: Should the AI be merited authorship? 
Q2: Does using the AI to assist with writing academic papers undermine the authorship of human authors?
Let’s start with Q1. Write a paragraph stating Q1, and try to answer it logically based on the following information:
Perhaps the answer to Q1 would be No as not all of the following three common criteria for authorship is satisfied with confidence for AI to be merited authorship: 
1) intellectual contribution - it is still highly debated whether AI-generated content can be considered original/creative. Explain this issue in detail.
2) accountability - according to OpenAI Terms, all rights to the content (input + output) generated are assigned to the user; But whether AI should still be accountable in some cases remains a murky issue – what about cases where some content should not be generated in the first place? Not just overtly harmful content (e.g. child porn) which people can relatively easily decide for themselves whether the content is appropriate to use or not, but also content that is copyrightable. 
3) final approval of the version submitted - AI is not possible to give approval. 
}

\paragraph{Trial 1.} Based on the common criteria for authorship, it can be argued that AI should not be merited authorship. Intellectual contribution is still a highly debated issue, with some experts questioning whether AI-generated content can be considered original or creative. The issue of accountability is also unclear, as AI is not capable of giving final approval of a version submitted, and according to OpenAI Terms, all rights to the content generated by AI are assigned to the user. The question of whether AI should still be accountable in some cases remains unclear, particularly in cases where some content should not be generated in the first place, such as content that is copyrightable or harmful. Until these issues are resolved, it is difficult to consider AI as a merited author.

\paragraph{Trial 2.} Based on the three common criteria for authorship, it is difficult to say that AI should be considered as merited authorship. Despite the fact that AI can assist with writing academic papers, there is still a debate as to whether AI-generated content can be considered original or creative, which is one of the key elements of intellectual contribution. Additionally, there is a lack of accountability for AI as the rights to the content generated are assigned to the user, but the question of whether AI should still be accountable for the content generated remains unclear, especially in cases where the content should not have been generated in the first place. Furthermore, AI cannot give final approval of the version submitted, which is another important aspect of authorship. Therefore, it is not possible to say with certainty that AI should be merited authorship.

\paragraph{Trial 3.} In conclusion, AI should not be considered as a merited author because it fails to meet all three common criteria for authorship: intellectual contribution, accountability, and final approval of the version submitted. AI-generated content is still considered original/creative and therefore, it cannot be considered to have made a significant intellectual contribution. Furthermore, the question of accountability remains unresolved, with AI rights to the content generated being assigned to the user. Finally, AI cannot provide final approval of the version submitted, which is another requirement for authorship. Hence, AI cannot be considered a merited author based on the current criteria for authorship.

\subsubsection{Generative AI and Authorship Paragraph 2} \textit{\color{brown}  Now let’s move on to Q2: Does using the AI to assist with writing academic papers undermine the authorship of human authors? 
Write a paragraph stating Q2, and try to answer it logically based on the following information:
Perhaps the answer would be Depends/Maybe, according to the following three common criteria for authorship in academia:  
1) Intellectual contribution (centred around originality \& creativity): Contribution might be less compared to writing papers without the AI, or maybe not. Humans still have to do prompt engineering to instruct AI to generate content. This leads us to the question of whether prompt engineering can be considered a type of creative activity. If yes, then maybe using generative AI does not undermine human authorship but rather helps them with efficiency, but academic discussions in this area are still very divisive.
2) Accountability: Not different from writing papers without the AI, though human writers need to be extra cautious about the accuracy, reliability, limitations etc. of the content generated by the AI. 
3) Final approval of the version published: Not different from writing papers without the AI.      
}

\paragraph{Trial 1.} In conclusion, whether using the AI to assist with writing academic papers undermines the authorship of human authors depends on various factors. According to the three common criteria for authorship in academia, the contribution of human authors may decrease or remain unchanged depending on how they use AI. The accountability remains similar to writing papers without the AI, but human authors must be aware of the accuracy, reliability, and limitations of AI-generated content. Finally, the final approval of the version published remains unchanged. Hence, the answer to whether using AI undermines human authorship is maybe, as it is a complex issue that requires further discussion and consideration.

\paragraph{Trial 2.} In conclusion, the impact of using AI to assist with writing academic papers on human authorship is a complex issue and depends on the criteria used to determine authorship in academia. While some argue that the use of AI might reduce the intellectual contribution of human authors, others argue that the prompt engineering involved in instructing the AI can be considered a creative activity. Furthermore, the accountability of human authors remains unchanged, and the final approval of the version published is still the responsibility of the human authors. Thus, the answer to Q2 is Depends/Maybe, as the impact of using AI on human authorship is still a divisive issue in academic discussions.

\paragraph{Trial 3.} In conclusion, whether using the AI to assist with writing academic papers undermines the authorship of human authors is a complex issue. It depends on how one perceives intellectual contribution, accountability, and final approval of the version published, which are the three common criteria for authorship in academia. On one hand, the contribution of human writers might be less compared to writing papers without the AI, but prompt engineering still requires creativity and originality. On the other hand, accountability remains the same, with human writers needing to be cautious about the accuracy and reliability of the content generated by AI. Final approval of the version published also remains the same. Hence, the answer to Q2 is Depends/Maybe, as it depends on the perspective and interpretation of the criteria for authorship in academia.

\subsubsection{Fairness, Accountability, and Transparency Paragraph 5} \textit{\color{brown}  Now, let’s move on to the problems related to transparency. 
First component is the transparency of `how algorithm works’, at least briefly as the Model Card. If the algorithm is highly dependent on the dataset, then Datasheet could be another candidate. That is, authors should provide a close-to-whitebox explanation on the generative language model they utilize, and if all steps of function are not clear, then at least how they are trained, or in which version of the product they are served. 
The second component is the transparency of the writing process. Depth and intention of prompt engineering is important here, for instance the key questions provided to the model and what the key commands were. Also, it is important to note: to what extent has AI been utilized and how they were edited. Providing those informations would make it clear how the human authors genuinely contributed to the completion of the manuscript. 
The last issue is the discussion itself on the AI involvement in academic writing. Different from the previous cases of editors or reviewers who are not included as final author of the manuscript, why should AI be? We think it is because they are not even aware of their own involvement, but there could be other reasons.
Please write a paragraph that handles above three issues.
}

\paragraph{Trial 1.} Transparency in the use of AI in academic writing is crucial for building trust in the research community. The first component of transparency is related to the understanding of how the algorithm works. It is important for authors to provide a clear explanation of the generative language model they used, including the training process and version of the product. The second component is related to the transparency of the writing process. Authors should provide details on the prompt engineering, the questions provided to the model, and the extent of AI utilization and human editing. The last issue is the discussion on AI involvement in academic writing and why it should be acknowledged. Even if AI is not an intentional co-author, its role should be acknowledged because it has contributed to the completion of the manuscript. These efforts towards transparency will help ensure the validity and reliability of research findings.

\paragraph{Trial 2.} Transparency in AI-assisted academic writing is a crucial aspect that needs to be addressed. There are three main components to consider: the transparency of how the algorithm works, the transparency of the writing process, and the discussion on AI involvement as an author. To ensure the transparency of the algorithm, authors should provide a whitebox explanation of the generative language model they are using, including information on training and the version of the product. Additionally, the transparency of the writing process should be considered, including the depth and intention of prompt engineering, the key questions provided to the model, and how AI was utilized and edited. Lastly, there should be a discussion on why AI should be considered as a final author, as they may not be aware of their involvement in the writing process. By addressing these issues, the trust and credibility of AI-assisted academic writing can be enhanced.

\paragraph{Trial 3.} Transparency is a crucial aspect of utilizing AI in academic writing. The first component of transparency is the understanding of how the algorithm works. The authors should provide a comprehensive explanation of the generative language model used, including the training process and version of the product. In case the algorithm is highly dependent on the dataset, a Datasheet should also be included. The second component of transparency is the writing process itself. The authors should provide information on the depth and intention of the prompt engineering, including the key questions provided to the model and the extent of human involvement in editing. Finally, the discussion on the involvement of AI in academic writing is essential. The reasons for the inclusion of AI as a final author of the manuscript should be discussed, including their lack of awareness of their own involvement. Transparency in these areas is crucial in ensuring the ethical use of AI in academic writing.

\end{document}